# Monitoring food structure during digestion using small-angle scattering and imaging techniques

Jade Pasquier[1,2,3], Annie Brûlet[2], Adeline Boire[4], Frédéric Jamme[3], Javier Perez[3], Thomas Bizien[3], Evelyne Lutton[1], and François Boué[1,2]

[1] *Génie et Microbiologie des Procédés Alimentaires, UMR0782 INRA-AgroParisTech, UPSay, F-78850 Thiverval-Grignon, France*
[2] *Laboratoire Léon Brillouin, UMR12 CEA-CNRS, UPSay, CEA-Saclay, F-91191 Gif-sur-Yvette, France*
[3] *Synchrotron SOLEIL, UPSay, F-91192 Gif-sur-Yvette, France*
[4] *Biopolymères, Interactions Assemblages, UR1268 INRA, F-44316 Nantes, France*

*Various studies have shown that food structure has an impact on digestion kinetics. We focus here on the effects of gastric and intestinal enzymes (in-vitro digestion) on two canola seed storage proteins, napin and cruciferin. To monitor structure effect we conducted experiments on gels of these proteins at different pHs, yielding different structures and elastic modulus. What is new is to get information on the mechanisms at the lowest scales, using imaging and radiation scattering at large facilities: Synchrotron fluorescence microscopy, X-Ray scattering, at SOLEIL synchrotron, and Small-Angle Neutron Scattering, at Laboratoire Léon Brillouin reactor. We can identify the mechanisms at each step and in two distinct scale ranges, observed simultaneously, the one of the individual protein scale and the one of the structure connectivity:*

- *during gelation individual canola proteins are not deeply modified in comparison with their state in solution ; larger scale gel heterogeneity appears due to connectivity or aggregation*
- *in the gastric step (up to 40 min):*
    - *at short scale (large q) we see that the proteins disintegration is much slowed down in gels than in solutions, particularly in the gastric phase;*
    - *at larger scales (low q), we see that the gel structure is also self-resistant to the action of the enzyme (pepsin).*

*- in the intestinal step, such kinetics differences hold until major disintegration after no more than 15 min.*

## 1. Introduction

World dietary needs are increasingly headed towards plant proteins since it becomes more and more obvious that feeding the world can be achieved only with limited amounts of animal proteins. Canola is considered as the 2$^{nd}$ source of oil seed in the world[1]. It contains up to 50% protein on a dry basis, has high nutritional values with a well-balanced amino acid composition, in particular glutamine, arginine and leucine[2]. Nevertheless, the meal also contains undesirable components like glucosinates and phenols, which have toxic effects on human health when ingested. An adapted extraction process is thus needed for collecting proteins of interest without degrading their properties.

Two storage proteins predominate in canola seeds, namely 2S albumin napin and 12S globulin cruciferin. Napin is a strongly basic protein with a pI around 11. This alkanality can be explained by a high amidation of its amino acids. Its primary structure is made of two polypeptide chains, a small one and a large one, joined together by two inter-chain disulfide bonds. Its molecular weight is rather low, between 12.5 and 14.5 kDa. Cruciferin represents 60% of the total protein content of canola seeds. It is a neutral protein (pI = 7-7.5), made of six sub-units with a high molecular weight around 300 kDa[3]. Each monomer is composed of two polypeptide chains (30 and 20 kDa), linked by a disulfide bond. Under extreme conditions as with a pH very far from its pI or in urea solutions, the protein is unstable and its subunits dissociate[4].

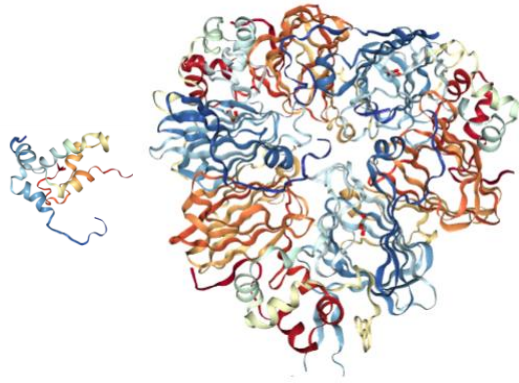

*Figure 1: representation of napin (left) and cruciferin (right). From Protein Data Bank.*

Leaning on experimental evidences that nutritional quality of food not only depends on its molecular composition but also on its structure, we have followed the degradation kinetics of these two canola proteins in the form of gel (compared to solutions) during in vitro digestion. Proteins gels are innumerable, and their structure can be various. It is now established that this structure has to be considered when evaluating nutrient absorption and possible health effects of food. Keeping whey protein as an example of very diverse structures[5], structure effects have been seen during digestion, in vitro[6], or in a human gastric simulator (HGS)[7]. The physical aspect of processes depending on the structure have been looked for, e.g. enzyme diffusivity, in these gels[8]. Digestion has been studied also in vivo, for other structures issued from dairy products: indeed a structure dependence has been established in the time dependence of leucine plasma content in minipigs[9] blood. There is thus a growing interest in the literature for this topic, which has extended to other kinds of proteins, such as egg albumin[10], and plant proteins[11].

There is also a strong need for new methods to collect information about the evolution of food micro and nanostructure during digestion. The work of Floury et al.[12] initiated the use of deep-UV fluorescence microscopy (DISCO beam line, Synchrotron SOLEIL), which will be used here. We will also look at smaller sizes (1 nm to 100 nm), which are accessible through scattering techniques: Small-Angle X-ray Scattering (SWING beam line, Synchrotron SOLEIL) and Small-Angle Neutron Scattering (PAXY, Laboratoire Léon Brillouin).

In brief, the objective here is to use these three techniques to monitor the microstructural changes induced by hydrochloric acid and digestive enzymes on two canola protein matrices, napin and cruciferin, from solution to gel state under in vitro digestion.

## 2. Materials and methods

### 2.1. Isolation of canola proteins

Cruciferin and napin were isolated from canola meal using a succession of chromatographic separations carried out at BIA laboratory, INRA, Nantes.

#### 2.1.1 Extraction

Canola meal was dissolved in a mixture of Tris 50 mM, NaCl 1 M, sodium disulfite 0,3% and EDTA 5 mM, at pH 8.5. After 1H at ambient temperature, the suspension was centrifuged on Rotor JSP F500 at 17000 RCF for 25min at 20°C. The supernatant was recovered and stored at 4°C. The pellet was re-extracted once again under same conditions. The two supernatants were then combined, vacuum filtered and the filtrate was stored at 4°C.

#### 2.1.2. Desalting separation

The aim was to separate canola proteins from polyphenols and salts. Through a Tris 50 mM, sodium disulfite 0,3%, and EDTA 5 mM buffer at pH 8.5, the filtrate was introduced into a size-exclusion chromatographic column fixed at 3,5 kDa (100 x 887 mm with cellulose gel Matrex cellufine GH25, 45-105 µm Amicon-Millipore). The first recovered fraction was composed of canola proteins. The process was stopped when conductivity increased, that corresponds to the arrival of salts.

A part of the mixture napin/cruciferin was set apart at this stage and dialyzed against osmosis water and ammonium carbonate during 48H with bath replacement every 8H. The resulting solution was then frozen and lyophilized. It contains a mixture of 29% napin, 35% cruciferin, and 36% residual molecular species, which is called "Mel" below.

### 2.1.3. Protein separation

Through a column SP-Sepharose fast flow (sulfopropyl cross-linked agarose 6% matrix, 90 µm, 300 mL XK 50/15 Amersham), napin and cruciferin were separated using cation-exchange chromatography. The mobile phase was a buffer made of Tris 50 mM and EDTA 5 mM, at pH 8.5.
Napin, with positive charge, was fixed on the resin whereas cruciferin was not held back and was directly eluted. Then, napin was recovered by addition of a buffer of Tris 50 mM, NaCl 1 M and EDTA 5 mM at pH 8.5, which has a stronger ionic concentration than the protein.

### 2.1.4. Purification of napin

The resulting napin was concentrated on a crossflow cassette Vivaflow 200 5 kDa Hydrosart Sartorius with a Tris 50 mM and NaCl 750 mM buffer at pH 8.5. Then, the protein solution was introduced on a gel filtration resin (Sephadex G50 medium 1,8L, XK50/100), with 1-30 kDa division, through a Tris 50 mM and NaCl 750 mM buffer at pH 8.5. This step consisted in separating napin from residual impurities according to UV absorbance at 280 nm. Napin fractions collected were combined and dialyzed against osmosis water during 48H with bath replacement every 8H. The solution was then frozen and lyophilized.

### 2.1.5. Purification of cruciferin

The purification of cruciferin was performed in BIA laboratory our coworker Alice Kermarrec, similarly to napin procedure.
Cruciferin was concentrated on a crossflow cassette Vivaflow 200 200 kDa Hydrosart Sartorius with a Tris 50 mM and NaCl 750 mM buffer at pH 8.5. Then, the protein solution was injected in a gel filtration column (Sephacryl 300 HR, XK 50/100), with 10-1500 kDa division, through a Tris 50 mM and NaCl 750 mM buffer at pH 8.5. This step consisted in separating cruciferin from residual impurities according to UV absorbance at 280 nm. Cruciferin fractions collected were combined and dialyzed against osmosis water and ammonium carbonate during 48H with bath replacement every 8H. The solution was then frozen and lyophilized.

### 2.1.6. Protein characterization

- **SDS–PAGE analyses** were performed for estimating the purity of napin and cruciferin after the protein separation step and at the end of the purification process, using 4-12% Gel Bis-Tris and MW SeeBlue Plus2 Prestained Standard.
- **N-Based Protein Content**: Protein percentage in purified powders, previously put in a desiccator 10 days before, was measured with a CNS Vario analyzer (Elemental) and a thermal conductivity detector for nitrogen and carbon.
- **ThermoGravimetric Analysis**: The humidity rate of all purified protein fractions was estimated using TG 2050, ThermoGravimetric Analyzer after 5 days in a desiccator. The proportions of each protein dry mass were deduced.

### 2.2. Solubilization and gelation of purified canola proteins

Three kinds of purified canola proteins were used: napin, cruciferin, and the mixture "Mel" of 29% napin/ 35% cruciferin/36% others, approximately.
Solubilization and gelation procedures were made according to C. Yang et al.[2] as follows. Powders were dispersed in a NaCl 0.5% solution with 0.1% of sodium azide (to avoid microbial development) in order to obtain a protein solution at 100 g/L. The suspensions were homogenized during one night at room temperature under gentle stirring. Then, their pH values were adjusted to 9 and 11 by introducing a few droplets of sodium hydroxide 1M and the protein solutions were centrifuged with an Ependorff MiniSpin Centrifuge at 10000 RPM, for 30min at 20°C. The supernatant was recovered and centrifuged once again under same conditions.

Then, solutions were transferred in sealed Luer Lock syringes of 2.5 mL and immersed in a water bath regulated at 95 C for 30 min. After gelation process, the syringes containing samples were cooled down at room temperature during 30 min and stored at 4 C. The visual aspect of a sample, of 5 mm diameter, was slightly turbid (not milky). This will be discussed below in relation with the SANS.

### 2.3. In vitro digestion

#### 2.3.1. In vitro gastric digestion kinetics

In vitro gastric digestion protocol was based on the standardized method of the research group Infogest, conducted by Minekus and coworkers[13]. The driving concept is to stay close to real human digestion while making protocols easy to use by a majority, hence enabling comparisons and debate. In the gastric step, pepsin, which is a protease, is the only enzyme introduced; it is admitted that the HCl/porcine pepsin mixture is satisfactory close to the composition of the human gastric juice. This is different for the intestinal juice; we use pancreatin from pigs, which contains lipases, proteases, amylases, etc. - many enzymes which are not really needed for the digestions of proteins and protein derived gels. However, since the gels contain only proteins, we consider that proteases only will act, the other enzymes being of no effect here (since enzymes are in weak concentration, they do not participate to the scattering). Meanwhile we stay within the INFOGEST protocol, which is very general: it can be used for many different kinds of food, and is the subject of debate and control in a very lively group of scientists.

We take the example of Mel solution: 235 μL of solution at pH11 were introduced in a test tube. 65 μL of HCl 1 M was added in order to reach the pH of 2-3 in the mixture. Then, 8 μL of porcine pepsin at 78000 U/mL (kept at -20 C since the delivery from the provider) was introduced to have 2000 U/mL in the final gastric solution. The tube was incubated in a water bath regulated at 37 C, like human body temperature, for 20 min. A second gastric digestion was performed in parallel for 40 min. At the end, samples were removed and positioned between two quartz (Kapton) plates in a cell to be analyzed by SANS (resp. SAXS).

The same procedure has been used for other solutions prepared at other pHs, see Table 1.

| | |
|---|---|
| **Mel solution at pH6** | **250 μL of sample** |
| | **500 μL of HCl 75mM** |
| | **18 μL of porcine pepsin at 84000 U/mL** |
| **Mel solution at pH9** | 250 μL of sample |
| | 500 μL of HCl 100mM |
| | 18 μL of porcine pepsin at 84000 U/mL |
| **Mel solution at pH11** | 235 μL of sample |
| | 65 μL of HCl 1M |
| | 8 μL of porcine pepsin at 78000 U/mL |

*Table 1: Composition of gastric mixtures for canola proteins solutions*

Only for solutions at pH11, a measurement was made without the addition of porcine pepsin (from pig stomach) in order to estimate the effect of HCl alone on the samples.

For gels, we detail first the example of the "Mel" gel at pH 11. 250 (±3.0) mg of Mel gel piece at pH11 were introduced in a test tube. 500 μL of HCl 100 mM was added in order to reach the pH of 2-3 in the mixture. Then, 18 μL of porcine pepsin at 84000 U/mL was introduced to have 2000 U/mL in the final gastric solution. The tube was immerged in a bath of ice for one hour in order to give enough time to the gastric solution to penetrate (permeate) inside the gel without activation of gastric enzyme. After one hour, the tube was incubated in a water bath regulated at 37 C for 30 min, in order to make the enzyme active through the whole gel. At the end, the sample was removed and put in a cell to be analyzed. It was then expected that the enzyme concentration inside the gel piece is homogenous.

The same procedure has been used for other gels; proportions are given in Table 2 for the "Mel" samples. Only for gels at pH11, a measurement was made without the addition of porcine pepsin in order to estimate the effect of HCl alone on the samples.

| | |
|---|---|
| **Mel gel at pH6** | **250 mg of sample** |
| | **500 µL of HCl 100mM** |
| | **18 µL of porcine pepsin at 84000 U/mL** |
| Mel gel at pH9 | 250 mg of sample |
| | 500 µL of HCl 150mM |
| | 18 µL of porcine pepsin at 84000 U/mL mg/mL |
| Mel gel at pH11 | 250 mg of sample |
| | 500 µL of HCl 150mM |
| | 18 µL of porcine pepsin at 84000 U/mL |

*Table 2: Composition of gastric mixtures for canola proteins gels. For the different pHs, the HCl volume has been adjusted to reach a pH of 2-3, similar to what happens in the stomach, the other volumes remaining the same (sample and pepsin).*

### 2.3.2. In vitro intestinal digestion kinetics

In vitro intestinal digestion protocol was also based on the standardized method of the Infogest group, on samples made out of gels or solutions prepared in the same way, and first subjected to gastric digestion.

200 µL of sodium bicarbonate 1M, 69.2 µL of porcine pancreatin (i.e. from pig) at 1600 U/mL and 69.2 µL of porcine bile at 160 mmol/L were added to the predigested mixtures of sample/gastric solution. The pH rose to 7, the optimal value for intestinal enzyme activation.

For gels only, tubes were then immerged in a bath of ice for one hour in order to let the intestinal solution permeate the gel without activation of intestinal enzyme.

All the tubes containing the intestinal mixture were finally incubated in a water bath regulated at 37 C. 5 min of intestinal digestion was achieved for Mel solutions and gels at pH 6 and 9. Another intestinal digestion was performed during 15 min for cruciferin and Mel gels at pH11, more resistant to intestinal enzyme action. At the end, samples were removed and placed between the two quartz (Kapton) plates of the cells to be analyzed by SANS (resp. SAXS).

### 2.4. UV X-ray microscopy

We used SOLEIL Synchrotron UV radiation on DISCO sent on the full-field microscope TELEMOS; (Excitation: 275 nm; Emission: DM300 327-353a) with a temperature-controlled plate regulated at 37°C, the human body temperature. A new injection cell, made of brass, for good heat conductivity, fitted to this plate, was designed enabling us to position the sample at reduced distance from the objective. Less than (100 x 100 x 100) micron$^3$ gels pieces could be kept in the beam in the center of the cell, thanks to a set of little columns (0.5 x 3 mm) arranged in a circle.

Under UV exposure, the protein samples fluorescence emission decreases initially fast (e.g. a life time of 5 minits), due to natural "bleaching". First we located a piece of gel of representative shape and aspect with the 10x objective, then with the 40x objective, and induced a "bleaching" with successive exposures of UV (50s) and visible (100ms). Second, using a peristaltic pump, the gastric solution (HCl 2.7 mM + pepsin 26 mg/mL) was gently poured on the gel pieces and the monitoring of the disintegration kinetics was started. Just after injection of gastric phase, we clearly observed the arrival of a "bright wave" moving through the aggregate (see section Results).

In the last phase, we injected the intestinal solution. We used a syringe-pump remote control, which enabled us to keep the microscope room in the dark, thus to obtain a clear t = 0 for this last phase.

### 2.5. Small-Angle Neutron Scattering

The measurements were done on the PAXY spectrometer at Laboratoire Léon Brillouin, CEA, Saclay. Three q-ranges were chosen, see table 3.

|          | Sample / Detector distance (m) | Wavelength (Å) | Counting Time (s) |
|----------|-------------------------------|----------------|-------------------|
| High q   | 1                             | 6              | 600               |
| Medium q | 3                             | 6              | 1800              |
| Low q    | 6.7                           | 15             | 3600              |

*Table 3: q-ranges for SANS settings.*

In order to keep the sample in a state of digestion, obtained previously in the adjacent chemical laboratory, the board hosting the cells in the neutron beam was regulated at 4 C and kept in dry nitrogen owing to a bonnet with quartz windows transparent to neutrons. Condensation on these windows was efficiently avoided thanks to an airflow sent on the board.

### 2.6. Small-Angle X-ray Scattering

#### 2.6.1. Small-Angle X-rays Scattering spectrometer

Data were collected on the SWING beam line at Synchrotron SOLEIL. Two distances sample/detector were used, i.e. 2m and 6m with a q range from $2 \cdot 10^{-3}$ Å$^{-1}$ to $2 \cdot 10^{-1}$ Å$^{-1}$. For some Figures, intensities were converted in absolute units using water as a reference sample.
For kinetics of gastric digestion, a vertical "board" (metal stage) was used to host two types of cells:
- aligned closed cells with mica windows, aligned on 4 horizontal rows, pre-filled with the samples, to analyze them either in their initial gelled form or predigested by a gastric solution during a given time.
- open cells, on the top row of the board, where liquids (gastric solution + sample) could be injected with a syringe, allowing a monitoring of gastric digestion kinetics, starting a few tenth of seconds after injection ("t = 0").
Circulating water in the "board" from a temperature controlled bath enabled regulation at 37°C.

#### 2.6.2. In vitro gastric digestion kinetics

We report below the protocol used for Mel solution and gels for X Rays, which is very close to the one used for neutrons.
**For solutions**, at pH 11, 236 μL of solution were introduced in a test tube. 66 μL of HCl 1 M was added in order to reach the pH of 2-3 in the mixture. Then, 7.8 μL of porcine pepsin at 78000 U/mL was introduced to have 2000 U/mL in the final gastric solution. The tube was put in ice before being placed on the device regulated at 37°C, to start the digestion kinetics on live.
**For gels**, after several tests, a gastric mixture was made for each Mel gel at pH 11. 62 mg of sample was put in a beaker, and 7800 μL of HCl $5.10^{-3}$ M, plus 200 μL of porcine pepsin at 78000 U/mL were added.
The beaker was immerged in a bath of ice for half an hour in order to let the gastric solution penetrate inside the gel without activation of gastric enzyme. Then, the beaker was incubated in a water bath regulated at 37 C for 120 min. At the end, the sample was removed and placed in a cell to be analyzed.

## 3. Results

### 3.1 UV X-ray microscopy imaging

Some images are presented below (an .avi video file is attached to the paper, see SI) of a preliminary experiment on a mixture of napin and cruciferin (29% of napin and 35% of cruciferin approximately). The pH conditions respect the Infogest protocol, hence are the same as for SANS and SAXS. Time t = 0 (Figure 2) represents the end of bleaching; at t = 5 min pepsin was introduced. An apparent "wave" of pepsin (not yet bleached hence strongly fluorescing) then started to interact with the gel piece, as seen formerly for dairy gels[12]. This was followed by a progressive degradation of the gel piece from 8 min to 13 min, at which intestinal juice was introduced. Then a fast and complete disintegration of the aggregate was noted. Finally an optical exploration of the whole inner content of the cell - i.e. of all gel pieces showed that no gel pieces were observable wherever.

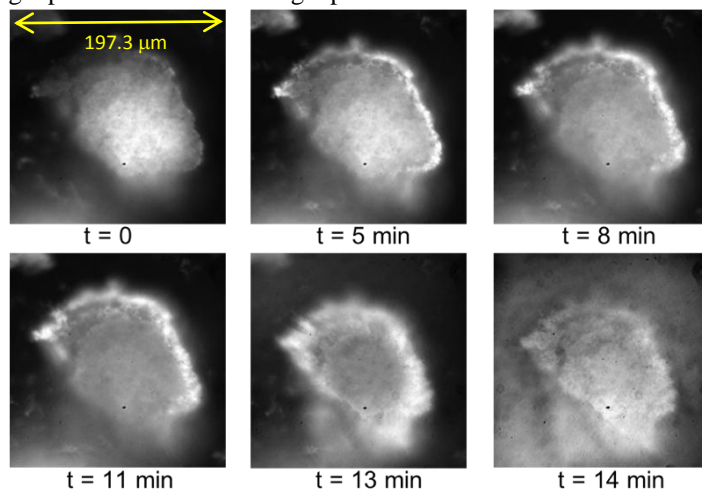

*Figure 2: Fluorescence imaging on the DISCO beam line, Soleil Synchrotron; image size: 137.93 x 137.93μm. Digestion of a gel of mixture (« Mel ») at pH 11, the gastric phase starts at t =5 minutes, the pepsin progressively diffuses inside the piece of gel (white area at the periphery. The intestinal juice has been injected at t =13 min, the piece of gel then dissolves quickly.*

These experiments are at a preliminary stage for plant proteins. They will be carried on, in order to understand the permeation of enzyme, following measurements on dairy gels[12]. They enable visualization in conditions close to real digestion conditions, i.e. when the food is ingested. This is complementary and will enable to build consistency with the very different ones used for SAS results, where the enzyme is first permeated at low temperature through the gel before starting digestion, which is necessary to bring a homogenous sample into the beam. At present, they provide a first information to build this consistency: they give a useful estimate of the digestion times.

## 3.2. Small angle scattering: initial state of solutions and gels.

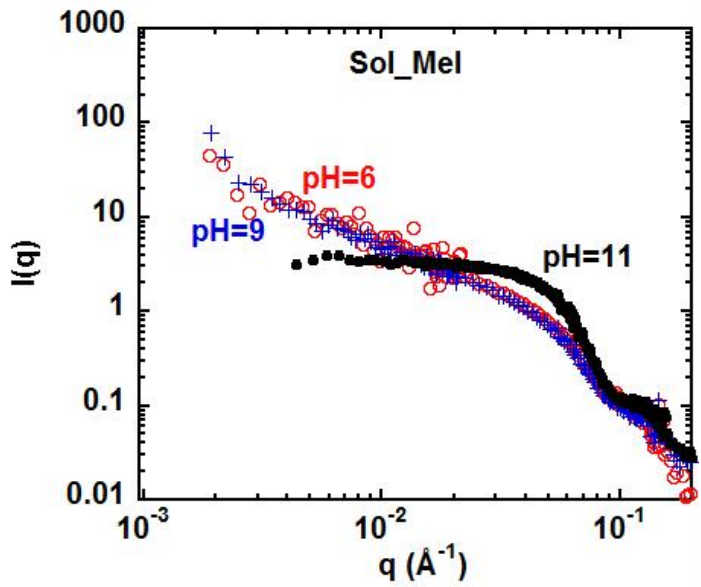

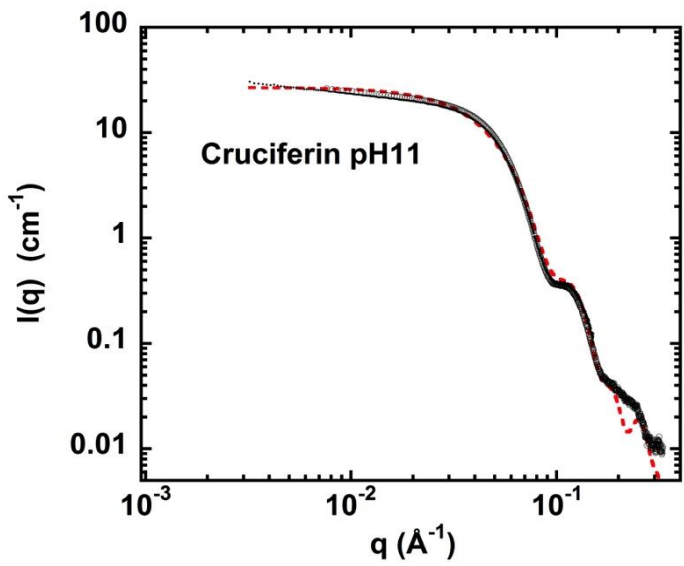

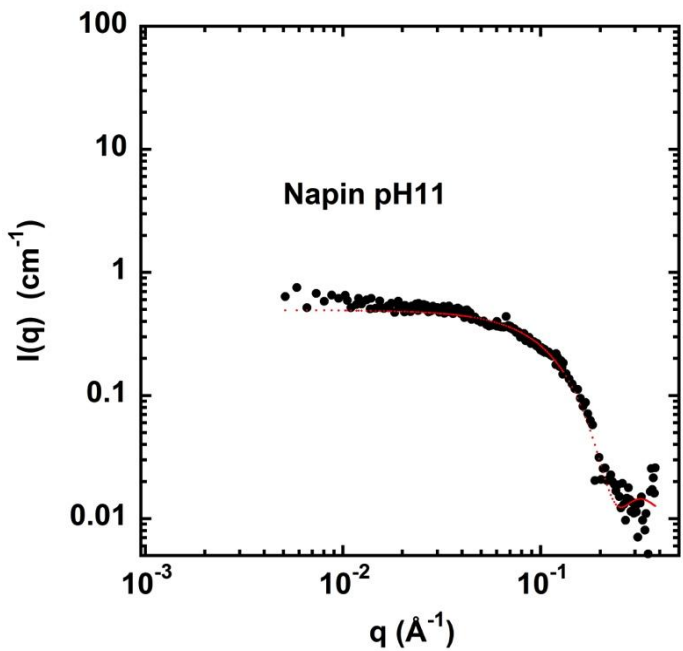

*Figure 3: Scattering of solutions in log I(q) versus log(q).*
*Top: SANS from the Mixture (cruciferin plus napin noted Mel, in $H_2O$) at the three pH at initial state (absolute units, $cm^{-1}$), PAXY, LLB*
*Middle: SAXS from 10% w/w cruciferin solution at pH11 – SWING, Synchrotron SOLEIL, with fit to a hollow cylinder of inner diameter $R_{core}$=10 +/- 1Å $\sigma$ =0.05, external radius R=46 +/-3Å, with polydispersity $\sigma_{ln}$=0.05 (lognormal distribution), length L=52 +/- 3Å.*
*Bottom: SANS from napin solution (in $H_2O$), pH 11 – PAXY, LLB, with fit to a compact sphere of radius R=18 +/- 2Å, $\sigma$ =0.05, concentration 22 g/L.*

Figure 3-Top shows the scattering of **solutions of the Mixture** ("Mel", cruciferin plus napine) at the three pHs for initial state.

At pH 11, the curve is very flat below $2 \cdot 10^{-2}$ Å$^{-1}$ ("plateau"), which is a sign that no aggregated proteins are present in this solution. Increasing q, a steep descent takes place above 0.04 Å$^{-1}$; such combination of ("plateau") and descent is called a shoulder and is characteristic of the protein form factor (compact shape). A second "shoulder" is visible at larger q = 0.1 Å$^{-1}$ (this shoulder is also seen using SAXS, see Figure S.I.5). Such shoulders are also observed on pure cruciferin (10g/L), as measured using SAXS (Figure 3 - Middle). They are related to a torus shape (sometimes the word "donut" is used) of the protein; the form factor has been calculated[14]. To fit our data, we have used a vicinal form factor, the one of a hollow cylinder – of simple access with SASview® (the torus form factor proposed by SASfit® will be used as soon as possible). Figure 3 - Bottom shows the scattering from the Napin solution (for 22 g/L), with a fit to a sphere form factor. The zero q limit is lower than 1 cm$^{-1}$ due to the small size of individual Napin and the absence of visible aggregates. We conclude that the scattering of cruciferin, which is much larger than napin (300 kDa compared to 14 kDa), is dominant in the mixtures ("Mel").

At lower pH, 9 and 6, the signal shows a slower decay at medium q: the bend of the shoulder is less steep. However the largest q behavior shows, like for pH11, a q$^{-4}$ average slope characteristic of compact objects. These two features are in favor of a slight, incomplete, unfolding of the proteins for the three different pHs. At low q, conversely to the case of pH 11, we observe at pH 9 and 6, a slight increase when q → 0, with apparent slope of order 2, i.e. a slight aggregation of some proteins due to change of interactions between some proteins after their slight unfolding.

**For gels** (we recall they are obtained by heating of the solutions without additive), Figure 4.a shows the scattering at different pHs.

At high q, for pH 11, direct comparison of the solution and the gel, in Figure 4.b, shows that the oscillations in the high q part for solutions, characteristic of a well defined conformation of the cruciferin (and not present at pH 6 and 9), have disappeared for the gels at pH 11, suggesting a slight unfolding during gelation. For pH 9 and 6, the signal, not showing oscillations even in the solutions scattering, is strikingly similar in the gels. Hence the protein conformation, slightly perturbed already in the solution is kept after gelation. The scatterings for pH 9 and 6, which were superimposing for solution, remain similar for gels.

At low q < 0.04 Å$^{-1}$, the same Figures 4.a and b show clearly an "upturn", namely an increase of I(q) when going from larger to lower q. The slope (exponent) of this increase is < 3 for pH11, and even larger, 3.5 – 4, for pH 9 and 6, i.e. the aggregates appear more compact.

In summary, with respect to solutions at same pH, the main difference in the gel state is a low q upturn of slope between − 3 and - 4. This is also seen using SAXS in Figure SI.1 in Supporting Information. Upturns in scattering are often encountered in the scattering of gels, and have been discussed in details formerly. They are explained by:
- either various kinds of heterogeneous spatial distribution[15,16,17] giving various slopes. For slopes larger than 1.6 and lower than 3, a fractal geometry is often invoked.
- or partial phase separation between regions poor (resp. rich) in polymer, or like here, proteins), leading to slope above 3, up to 4. [18,19]

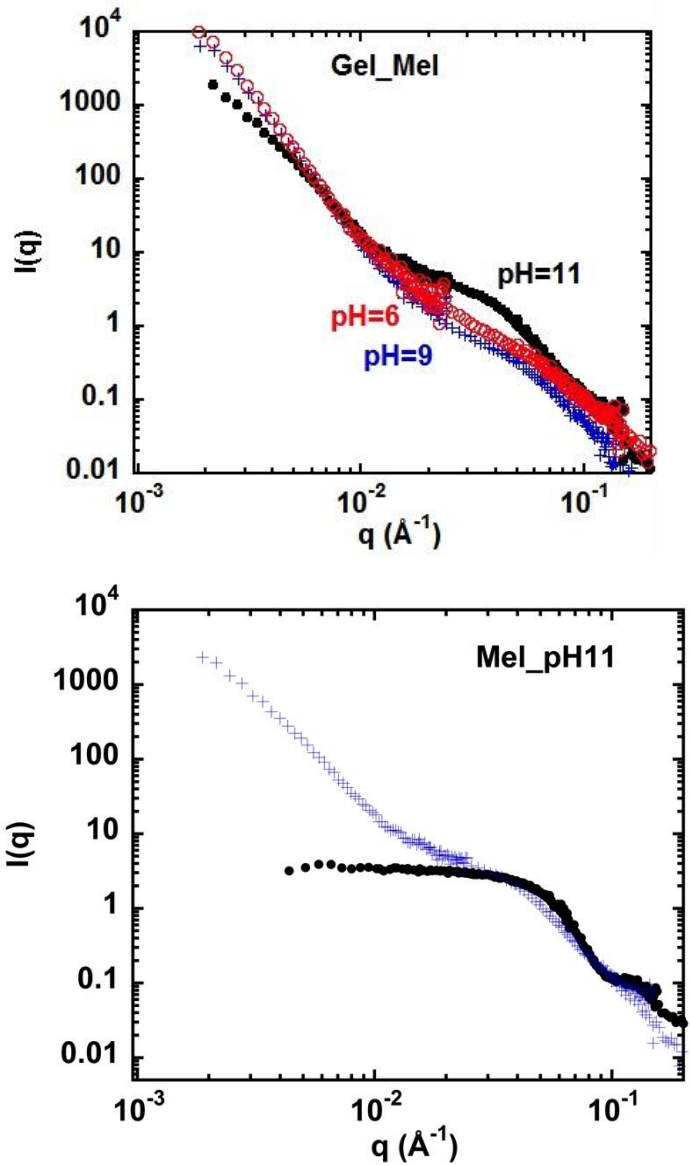

*Figure 4: SANS (a) above, gels obtained by heating from the cruciferin plus napin mixtures at the three pHs at initial state. (b)below, comparison gel / solution at pH 11. There can be a small difference of concentration with solutions, so we di not use the $cm^{-1}$ units.*

### 3.3 SAS: gastric and intestinal phase.

We first give a visual description of the evolution of the scattering, with a simple physical interpenetration in sections 3.3.1 and 3.3.2. On the Figures shown for this are also drawn solid lines, corresponding to fits described and commented in section 3.3.3 in a more synthetic manner.

### 3.3.1 The case of solutions.

**Sol Mel pH11.** When hydrochloric acid is added, the low q scattering increases strongly with a slope 4 (Figure 5). At intermediate q, the shoulder at 0.04 Å$^{-1}$ has also disappeared, which rubs out the scattering of individual proteins. This shows that most of the proteins have been aggregated because of this sudden change of pH. The low q scattering of the aggregates formed tends to a $q^{-4}$ Porod since they involve most of the proteins and are compact. This origin is slightly different from the one for low q scattering of the gels.

Then, after addition of pepsin, we obtain the final gastric mixture (noted HP). A reduction of intensity at low q has occurred after 20 min of gastric digestion and carries on even after 40 min (not shown here). The low q scattering from the aggregates formed under the effect of the acid decreased by 5.

Meanwhile, at large q, a plateau re-appears, suggesting that some individual proteins are removed from the aggregates. The intensity of this plateau, therefore the individual protein content is ten lower (which overpasses by far the change of concentration by dilution), which suggests that, since the aggregate signal has been strongly reduced that a significant fraction of the released proteins have been degraded.

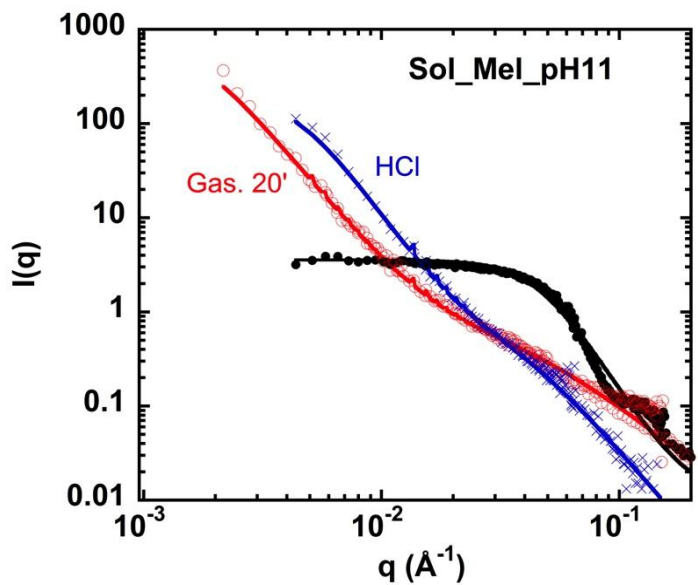

*Figure 5: SANS spectra for solution of napin/cruciferin mixture at pH11 (Sol_Mel_pH11), in log I(q) versus log q. Filled circle, initial state; cross, after addition of HCl; open circle, after 20min of action of pepsin with HCl. Note that the solution is diluted by the addition of HCl or gastric juice (75 µL added to 235µl, see Materials and Methods): the concentration has changed.*

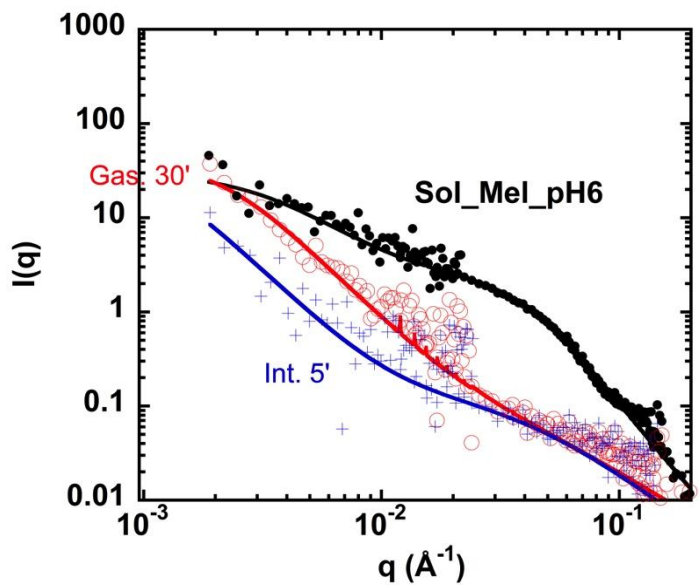

*Figure 6: SANS spectra with fits for solution of napin/cruciferin mixture at pH6 (Sol_Mel_pH6), in log I(q) versus log q (absolute units, $cm^{-1}$) – PAXY, LLB. Filled circles: initial state; empty circles: after 20min of action of pepsin with HCl; +: after 5 minutes of action of intestinal juice. Note that the solutions are diluted by the addition of HCl or gastric juice ((518 µL added to 250µl,): the concentration has changed.*

**Sol Mel pH 9 and 6.** Figure 6 and Figure S.I.1 and S.I.2 show the evolution of mixture ("Mel") solutions prepared at pH different from pH 11, respectively 6 and 9. Strikingly the curves for pH 6 and 9 are very similar at each step, which is a strong support of the validity of our experimental approach. For both, after 30 min of gastric digestion, intensity evolves qualitatively the same as at pH 11, but quantitatively more strongly (beyond the dilution effect by a factor 3): the resistance to gastric juice is weaker, probably because the structure is already affected. Finally, after intestinal digestion (5 min only!), aggregates go on being degraded faster than individual proteins.

### 3.3.2 The case of gels.

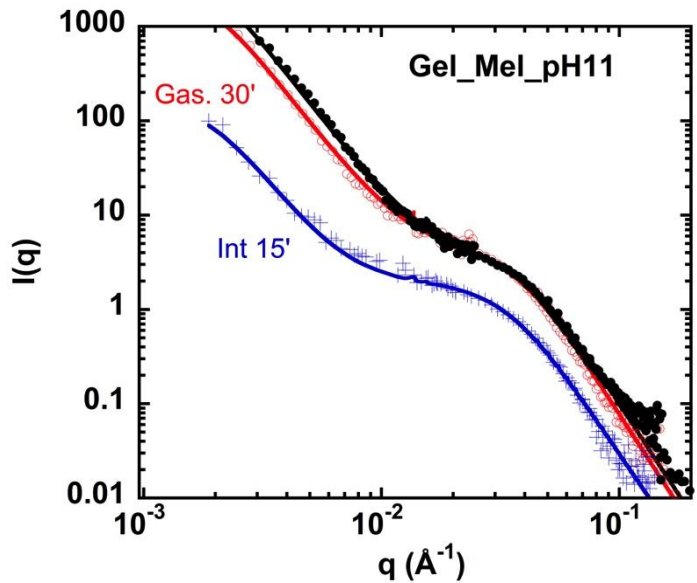

*Figure 7: SANS spectra for gels of napin/cruciferin mixture at pH 11 (Gel_Mel_pH 11), in log I(q) (arbitrary units: for gels there can be a small or large (intestinal step) uncertainty on protein concentration, so we did not use the $cm^{-1}$ units) versus log q.*

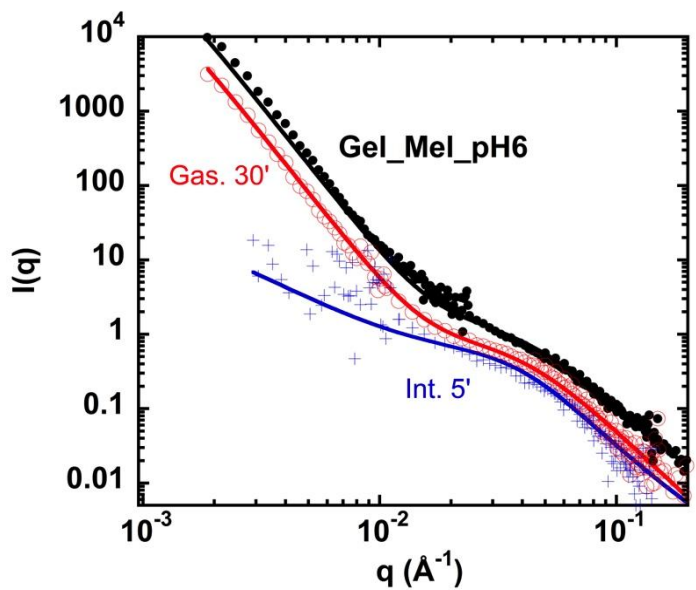

*Figure 8: SANS spectra for gels of napin/cruciferin mixture at pH 6 (Gel_Mel_pH6), in log I(q) (arbitrary units: for gels there can be a small or large (intestinal step) uncertainty on protein concentration, so we did not use the $cm^{-1}$ units) versus log q.*

For pH 11, Figure 7 shows the evolution of the scattering of gels, under digestion. After the action of HCl, the scattering (not shown here) is hardly changed: it superimposes with the initial one. After 30 min of gastric digestion, few changes have occurred: the intensity has a little decreased at low q, not at high q. This suggests that the gel has not swollen, and that proteins are not degraded at this stage, while a few aggregates are. This trend is much reinforced after only 15 min of intestinal digestion. The signal from proteins seen individually at large q is reduced by a factor around two. Note that, due to our protocol, there is some uncertainty on the volume of gel present in the cell (loss of material, swelling not excludable).

For pH 6, Figure 8, the same trend is visible, with differences (i) the whole intensity has decreased by two after the gastric step; (ii) after 5 min only of intestinal digestion, the low q intensity of diffusion is now very low. This can be disconnected from possible loss of sample in the cell or swelling, since the high q scattering stays the same as for the gastric step

Spectra for pH 9 are very similar to the one for pH6 (see Figure S.I.3 and Figure S.I.4), except at low q after 5min of intestinal digestion which appears much more efficient than for pH 6.

All together, it is clear that the structure of these gels makes proteins more resistant to digestive enzymes than in solution.

### 3.3.3 Fitting results.

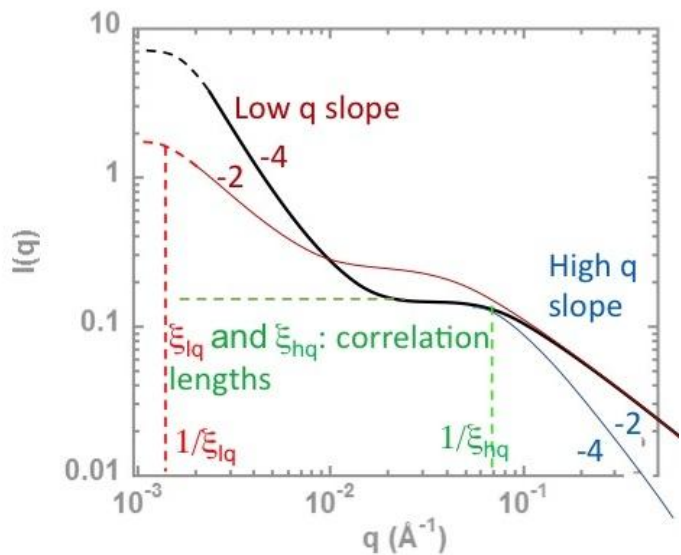

*Figure 9: Visualization of scattering parameters for fitting by Eq. 1.*

The former description being rather detailed, we now propose a modeling of data by fitting, trying to deliver more simple but still complete enough, information. It is obvious from the former description that the scattering always displays a partition between two components, in two different q regions. This is schematized on Figure 9, which evidences our fitting strategy: using the "Two Correlation length" model of Sasview[20], Eq. 1 introduces two terms of same mathematical form: one at low q describing the aggregation level of solutions or gels, and one at large q, describing the proteins seen individually:

$$I(q) = \frac{B_{lq}}{1+(q\xi lq)^{elq}} + \frac{B_{hq}}{1+(q\xi hq)^{ehq}} + Bk \quad (Eq. 1)$$

A background Bk, due to possible uncertainties in subtraction of the incoherent signal, can be added. Best-fit parameters are given in Table 4. The parameters $elq$, $ehq$, $\xi_{lq}$, $\xi_{hq}$, ▯▯▯ ▯▯▯ two intensity ▯▯▯▯▯▯▯▯▯▯▯▯ computed▯for the two terms of Eq. 1 ▯▯▯q = ▯▯▯04▯Å$^{-1}$ for low q and q =▯q = ▯▯▯5▯Å$^{-1}$ for high q are presented in a "histogram-like" representation below (Figure 10 and 11, a, b, c). We recall that the content in gel inside the neutron cell is not perfectly controlled. Indeed, we recovered the biggest gel pieces from the digestion test tube, but some pieces may have been separated from the larger ones and not recovered. So part of the decrease may be not due to digestion. This acts on the front factor of the scattering, id est on $B_{lq}$ and $B_{lq}$, and on □□□□□□□□□□□□□□□□□□ together. But this front factor should act neither on parameters $e_{lq}$, $e_{hq}$, $\xi_{lq}$, $\xi_{hq}$ nor on the ratio between the high and low q□□□□□□□□□□□□□□.

**High q fitting parameters**
**High q correlation length**. For solutions before digestion, the scattering is mostly related to the cruciferin form factor, and therefore not accounted by a simple correlation length. The fitted value s nevertheless not far from the correlation length, $\xi_{hq}$, obtained after gastric digestion (red), for which it better applies, since the protein form factor is slightly degraded: an other signature of degradation is that the bending spreads over a wider q range. All together, on Figure 10.a, at first sight, $\xi_{hq}$ does not vary much. There are two noticeable changes (i) for solution prepared at pH 11, a variation from 22 Å to 27 Å is noticeable for solution, due to pH change in the gastric juice; (ii) also in gels, after 40 min of gastric step, a variation from 25 Å to 33 Å even 40 Å, which suggests unfolding, but a rather limited one. At pH 6 the values are just the same for solution and gel, and do not vary during digestion. This suggests that the effective size of individual proteins still present is not changed.

In summary, the size associated to individual protein is increased by 30%, at pH 11 only (not for pH 6 neither for pH 9), probably due to the larger pH change during gastric digestion, or more precisely to crossing the isoelectric point region.

**High q exponent.** Figure 10.b shows that $e_{hq}$ decreases clearly under digestion for solutions. Apart from the case of immersion in HCl, we attribute this to some unfolding of proteins. The decrease is from 4 (compact) to 2 (random walk) in solution. For gel pH11 the unfolding seems limited (exponent $e_{hq} > 3.5$), except for 40 min gastric step (the one who shows an increase of $\xi_{hq}$). The slope even increases a bit for gels made at pH 6.

In summary, proteins still contributing to the scattering appear unfolded by digestion in solutions, but protected from the unfolding when in gels. This is observed both in gastric and in intestinal juice, even if the scattering is noticeably abated in the latter case, as we will describe just below.

**High q contribution.** First remember we are in log scale: this helps to visualize directly the correspondence with the scattering plots which are also in log scale, but we must keep in mind that we can have some decrease of the intensity by more than a factor ten!

The high q contribution (Figure 10 c) represents the scattering of individual proteins, for different states of conformation, as long as they are not compactly aggregated. This is why for solutions at pH 11 the intensity decreases fast after aggregation upon HCl, and re-increases afterward in gastric juice because some aggregates are partly "de-compacted" (expanded).

So for solutions, after gastric and intestinal steps, the effect of digestion is very clear because proteins have been cut into smaller sequences. For gels, intensity decrease is progressive for pH 11, and faster for pH 6 (as well as pH 9, not shown), but it is here pretty clear that the decrease is also slowed down.

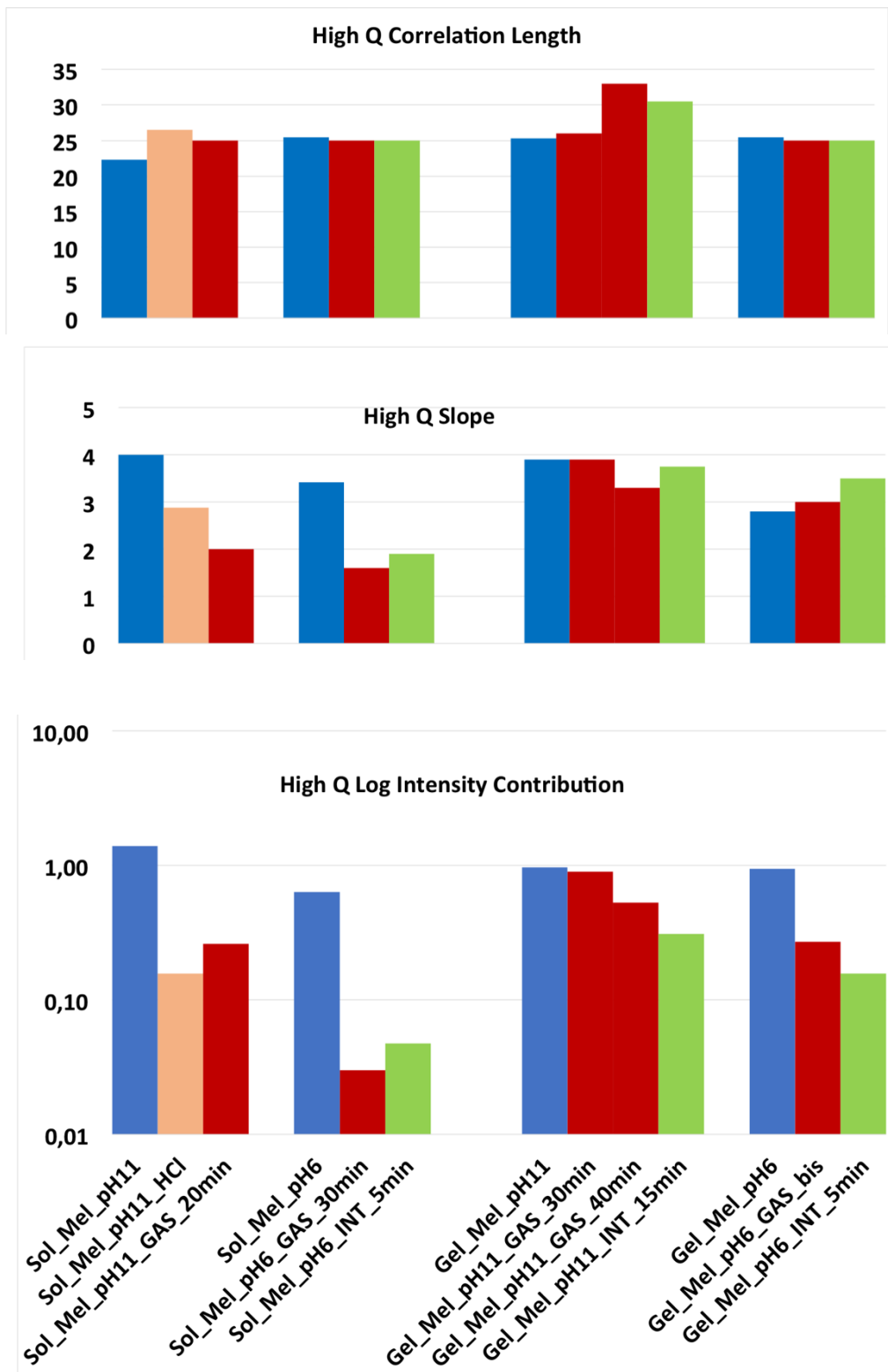

*Figure 10: Evolution of (a) High q correlation length, (b) High q exponent and (c) High q contribution, **in logarithmic scale** for solutions/gels of Mel at pH11 and 6, at initial state, after 20/30 min of gastric digestion, and finally after 5 or 15 min of intestinal digestion.*

**Low q fitting parameters**.

**Low q exponent.** For solutions: at pH 11, the exponent is zero initially, and jumps to 3 under the effect of low pH (HCl). At pH 6 (and 9, not shown), the low q exponent is initially around 2, and is neither sensitive to low pH nor to digestive enzymes. The corresponding non compact (open) structure resists digestion.
For gel, initial state, it is 3, or even 4 for pH6, due this time to the gel structure. The exponent resists gastric digestion; under intestinal digestion only, the gel aggregates or structure decrease; but it is strongly dismantled only at pH6.

**Low q correlation length.** For solutions, we observe a strong variation of $\xi_{lq}$ up to 500 to 1000 Å, since adding HCl only, or the presence of HCl in the digestion juice, have induced aggregation. For gels, large values are reached already in the initial state; they are kept under digestion, but data show low sensitivity in the explored q range, where $q.\xi_{lq}$ is too large ($> 1$).

**Low q intensity contribution.** The evolution is strikingly different for solutions and gels. In summary:
- solutions are quite sensitive to pH change, form aggregates, which vanish fast under digestion.
- conversely, in gels, aggregates are present initially, they resist HCl and gastric phase, while they noticeably dissolve after the intestinal one (5 min only for pH 6).

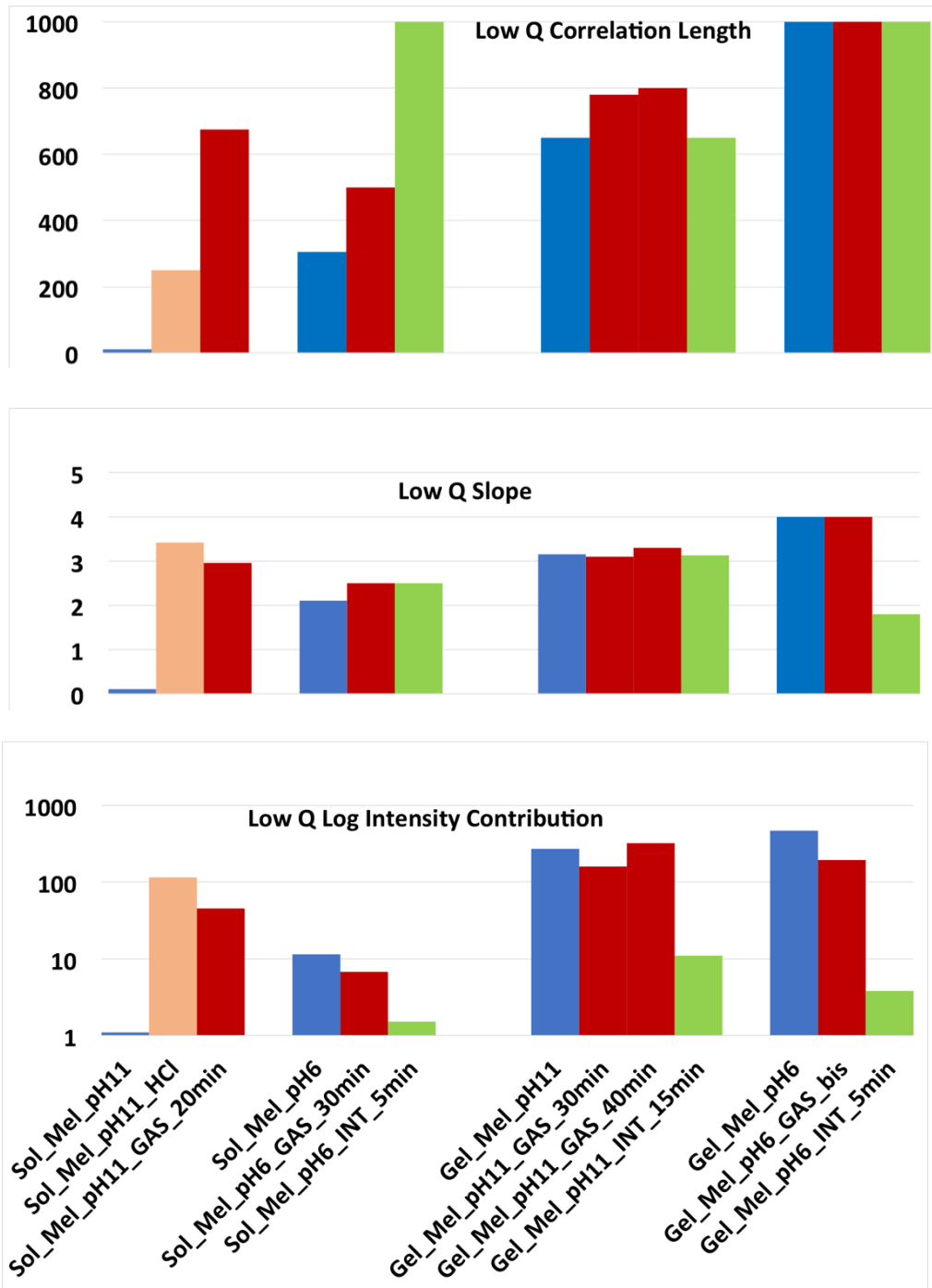

*Figure 11: Evolution of (a) Low q correlation length (b) Low q exponent and (c) "Low q intensity contribution", **in logarithmic scale,** for solutions/gels of Mel at pH11, and pH6, at initial state, after 20/30 min of gastric digestion, and finally after 5 or 15 min of intestinal digestion.*

### 3.4 Small-Angle X-ray Scattering

Before digestion, we observe by SAXS the same behavior than for SANS, as shown in Figure S.I. 5 for comparison of Mel sol and gel scattering. After digestion, Figure 12 shows the evolution of SAXS from a gel of Mel at pH 11. Compared to the case of SANS at only 30 min of digestion in Figure 7, the SAXS intensity has much more decreased after 120 min of gastric digestion (it is close to what observed after 15 min intestinal digestion). For $q > 5 \ 10^{-2}$ Å$^{-1}$ we observe the same curve shape with a vertical shift, which suggests a reduction of the quantity of sample in the beam. Meanwhile, at lower q, we observe a reduction of the apparent slope. So SAXS results agree with SANS results. A paper exploring the different potential (mainly kinetics) of SAXS is on the way.

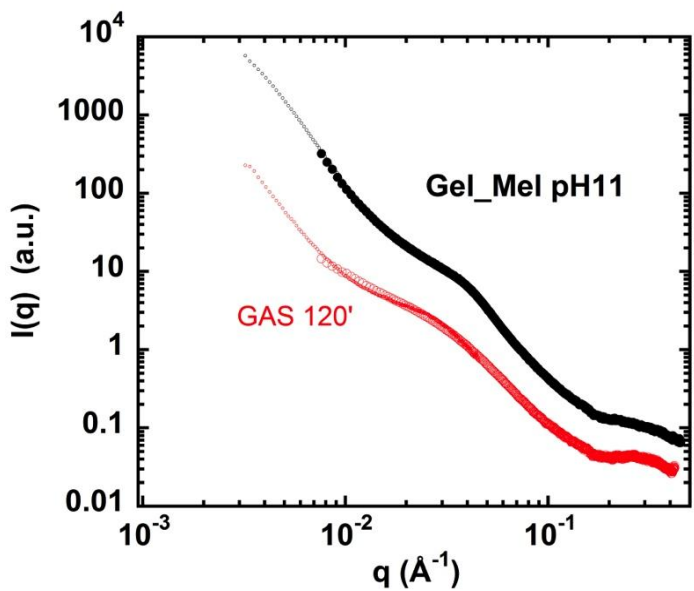

*Figure 12: SAXS spectra for gel of napin/cruciferin mixture at pH 11, in initial state and after 120 min of gastric digestion at 37 C in log I(q) versus log q (arbitrary units$^s$) – SWING, Synchrotron SOLEIL.*

### 4. Summary and discussion

#### 4.1 Initial state: well-defined scattering for an easy monitoring.

To study the influence of structure on digestion in this paper, we have first considered in vitro gastric step by monitoring the action of the dominant gastric enzyme, pepsin, which is a protease. We have therefore studied protein food, and then a simple approach for us was to compare solution and gels. Using purified simple compact proteins has enabled us to identify the starting point of our study, namely the signal of the solution of such compact objects. For some practical and cost reasons, we used a mixture of napin, quite small, and cruciferin, much bigger. The scattering is therefore dominated by the cruciferin. It appears very well dispersed in solution at genuine pH (pH 11), with a moderate additional signal from aggregates when pH is lowered to 9 or 6 (i.e. crossing the isoelectric points). The scattering can be fitted to the one of a hollow cylinder of well-defined radius, close to a torus, as known for native cruciferin. The starting solution is therefore well defined.

For gelation we have chosen heating, which does not introduce any new components in the system. There are obviously plenty of different proteins gels, and many structures have been extensively studied in the past. It is not the purpose of this paper to provide yet another such gel study. It is also obvious that other structures have to be studied under digestion to get more generality. Our purpose was to synthesize a structure, the evolution of which is easy to describe. This goal is nicely reached due to the presence of two well-separated q ranges. Namely:
- the high q range displays only the scattering of the proteins, very close from the solution one;

- the low q range is dominated by structure, aggregated or ramified, resulting from the heating. The large slope (3 to 4) corresponds to a rather strong aggregation, close to a phase separation.
- It is possible that the slight turbidity (i.e. light scattering) of the gels at visual inspection corresponds to the extension of the same signal in the lower q regime associated with the large wavelengths of light.

Two cases are possible:
- either most of the proteins are part of very large aggregates; in such case the aggregates cannot be compact (proteins in close contact with each other) because their scattering at high q would not show the protein one, which is observed. Hence the inside of these aggregates must be open – not compact. This would better agree with lower values of the low q exponent.
- or only a limited fraction $\phi_{InAgg}$ of proteins pertains to the aggregates, which can be either compact or open. This is more likely: meanwhile, a very large majority of the proteins stay, unaltered in initial state, in the gel.

Modeling the scattering of the aggregates would be simple, assuming a size and a compacity, if $\phi_{InAgg}$ was known, unfortunately this is not the case.

## 4.2 Information from the scattering during digestion.

When gastric juice is added two chemical processes run in parallel. The first one is just due to the lowering of the pH, then digestion by pepsin takes place.

### *4.2.1 Effect of HCl.*
We tested the effect of HCl alone:
- for solutions HCl generates an immediate aggregation, a very common effect, usually because the electrostatic repulsions are strongly perturbed, such that the solution is destabilized. The scattering at low q is hugely increased and the large q is abated, since aggregated proteins are not "seen" anymore as individuals. As expectable, the solutions at pH11 appear more affected by HCl addition than the solutions at pH6 (or 9), probably because the isoelectric point has been crossed only for pH11.
- for gels, conversely, there is hardly any changes. At low q, this could be masked by the large slope which is present initially before HCl addition. However, the scattering remains perfectly superposed to the initial gel one, as if its structure was unchanged. This is confirmed by the high q scattering, also unchanged, contrary to the case of the solution.

**Hence a first evidence of protecting effect of the gel structure appears here:** protecting for the proteins (high q), as well as self-protecting (low q).

### *4.2.2 Effect of Digestive juices.*
When the gastric juice, and then the intestinal juice is added:
- solution scattering evolves fast in both q ranges: the aggregated structure due to low pH is dismantled: it becomes less compact (decreased low q slope), and with lower intensity. High q slope decrease suggests that the proteins become more unfolded. Their individual contribution decreases noticeably in logarithmic scale (of order 1), i.e. rather fast for the actual value. E.g., for pH6 after arrival of the intestinal juice pushes the intensity has decreased by a factor ten, which suggests important protein digestion (even though dilution has to be after accounted for).
- gel scattering evolves slower than solutions: the shape of proteins is kept unchanged (as for HCl addition) much longer. The aggregate scattering decreases, but more slowly than solutions. There is no sign of swelling. In short, we observe here also a **delaying effect of the gel**: the gel structure slows down the action of the enzyme, as well as the one of HCl. This protecting effect extends to intestinal digestion. However, the progress degradation is obvious especially after within 5 min, for pH 6). At the macroscopic level, visually, gels are usually completely dissolved after 30 min at 37 C.

- getting to the comparison between gels at pH 11 and pH 6, the obvious effect is that degradation is faster at pH 6. This maybe explained by the stronger heterogeneity (aggregation), observed at pH 6, but also by the stronger unfolding of proteins.
- for solutions, aggregates (in the low q range) tend to dismantle as fast as individual proteins at the individual scale. This may due to the fact that they are made of partially unfolded proteins. In gels they are more resistant.

5. Conclusion

Through this study, using large instruments at Synchrotron SOLEIL and Laboratoire Léon Brillouin, our aim was to get a first approach of the disintegration of microstructured canola protein matrices in simulated digestion conditions. This project was based on cruciferin and napin, two canola storage proteins already studied in BIA-Nantes (INRA), and which own a higher degree of internal motion as compared to animal proteins, which may provide some regions more accessible for digestive enzymes. In this paper we showed results for pure proteins in their raw state, but restricted ourselves, for digestion studies, to a mixture in natural proportions. We compared solutions and gels formed under heating, not far from domestic or industrial cooking conditions.

Using deep UV fluorescence microscopy, preliminary experiments, in the line of the one achieved on dairy gels[12] showed that it is possible to follow enzyme permeation in vitro under conditions close to in vivo ones.

Using Small Angle Scattering, more achieved experiments focused on two kinds of sample structure, in the form of solution or gel, at different pH. Using these purified proteins, a moderate heating enables us to obtain a structure the scattering of which is simple to analyze. We observe a clear and useful separation between:

- the low q range, where the gel scattering is quite different from the solution one
- and the large q range, which enables us to follow the structure of the proteins at the scale of one entity.

The main effect is that proteins are not deeply modified during gelation, and are more protected from digestion once the gel structure is formed, compared to solutions. The aggregates are digested as fast as the proteins in solutions (possibly because the aggregated proteins are partly degraded already). In gels, they benefit also from self-protection of the gel. To make a cruder analysis of the results, we could say that the gel structure is not profoundly modified during digestion, even when the intensity has strongly decreased (intestinal digestion). It therefore suggests, within the experimental accuracy, that some parts of the gel remain unperturbed while the others are completely digested.

Beyond the possibility of a simple but quite consistent analysis, this study has, of course, some weaknesses. One is the use of a mixture of two proteins, quite different in molar mass. The analysis of the results is thus to be enriched by measurements on separate napin and cruciferin solutions and gels (to publish). A second one is the exact origin of the initial low q scattering. Is there a relation with the fact that two proteins are mixed? Or, are they separate aggregates, or embedded in the structure? Beyond more extended modeling with further scattering data, rheology measurements are now on the way to provide answers to these questions. But, to finish, a deeper question concerns the relation with real conditions of digestion inside the Gastro-Intestinal Tractus. For neutron scattering measurements, in order to have homogenous samples over a beam size of 10 mm, we have let the enzyme permeate the gel before to heat up to the enzyme working temperature 37 C. It is only using UV microscopy that preliminary studies enabled us to start to tackle the problem in situ. We intend to continue those studies with microscopy, but also we have undergone experiment with thinner beam to follow the scattering in similar conditions. The SANS experiments shown here establish a strong basis to analyze the corresponding measurements.

**Acknowledgements**
We thank LLB for beam time allocation on PAXY, with the great help of Arnaud Hélary. We thank SOLEIL synchrotron for beam time allocation on SWING and DISCO lines, with the great help of Youssef Liatimi, and Valérie Rouam.

We thank Thomas Cattenoz for his kind help at GMPA, and all GMPA and ADP team for general support.
We thank INRA (CEPIA Division), SOLEIL and LLB for the PhD grant of Jade Pasquier.
We thank BIA for hosting Jade Pasquier for an internship dedicated to the preparation of the proteins, with the great help of Alice Kermarrec.

*Table 4: Best fit parameters (Eq.1) of solutions and gels of mixture, before and after gastric / intestinal digestions.*

|  | Low q (Lorentz) | | | | High q (Lorentz) | | | |
|---|---|---|---|---|---|---|---|---|
|  | Correlation length $x_{lq}$(Å) | Exponent $e_{lq}$ | B_lq | Low q Intensity contribution at q=0.004Å$^{-1}$ | Correlation length $x_{hq}$(Å) | Exponent $e_{hq}$ | B_hq | High q Intensity contribution at q=0.04Å$^{-1}$ |
| Sol_Mel_pH11 |  | 0,1 |  |  | 22,3 | 4,0 | 3,6 | 1,4 |
| Sol_Mel_pH11_HCl | 250 | 3,42 | 231 | 115,5 | 26,5 | 2,9 | 0,5 | 0,2 |
| Sol_Mel_pH11_GAS_20min | 675 | 2,96 | 905 | 45,4 | 25,0 | 2,0 | 0,7 | 0,3 |
| Sol_Mel_pH6 | 305 | 2,11 | 29 | 11,3 | 25,5 | 3,4 | 2,1 | 0,6 |
| Sol_Mel_pH6_GAS_30min | 500 | 2,50 | 45 | 6,8 | 25,0 | 1,6 | 0,1 | 0,0 |
| Sol_Mel_pH6_INT_5min | 1000 | 2,50 | 50 | 1,5 | 25,0 | 1,9 | 0,1 | 0,1 |
| Gel_Mel_pH11 | 650 | 3,16 | 5891 | 274,0 | 25,3 | 3,9 | 3,4 | 1,0 |
| Gel_Mel_pH11_GAS_30min | 780 | 3,10 | 5600 | 159,8 | 26,0 | 3,9 | 3,4 | 0,9 |
| Gel_Mel_pH11_GAS_40min | 800 | 3,30 | 9509 | 323,0 | 33,0 | 3,7 | 5,1 | 0,5 |
| Gel_Mel_pH11_INT_15min | 650 | 3,13 | 223 | 10,7 | 30,5 | 3,8 | 1,9 | 0,3 |
| Gel_Mel_pH6 | 1000 | 4,00 | 120000 | 466,0 | 25,5 | 2,8 | 2,8 | 0,9 |
| Gel_Mel_pH6_GAS_bis_30min | 1000 | 4,00 | 50000 | 194,0 | 25,0 | 3,0 | 0,8 | 0,3 |
| Gel_Mel_pH6_INT_5min | 1000 | 1,80 | 50 | 3,8 | 25,0 | 3,5 | 0,5 | 0,2 |

**References**

[1] Aider M., Barbana C. Canola proteins: composition, extraction, functional properties, bioactivity, applications as a food ingredient and allergenicity – a practical and critical review, 2011, Food Science & Technology, 22, 21-39.

[2] Yang C., Wang Y., Vasanthan T., Chen L. Impacts of pH and heating temperature on formation mechanisms and properties of thermally induced canola protein gels, 2014, Food hydrocolloids, 40, 225-236.

[3] Perera S. P., McIntosh T. C., Wanasundara J. P. D. Structural properties of cruciferin and napin of Brassica napus (Canola) show distinct responses to changes in pH and temperature, 2016, Plants, 5, 36

[4] Bérot S., Compoint J. P., Larré C., Malabat C., Guéguen J. Large scale purification of rapeseed proteins (Brassica napus L.), 2005, Journal of Chromatography B, 818, 35-42.

[5] Nicolai, T., Britten, M., & Schmitt, C. Beta-lactoglobulin and WPI aggregates: Formation, structure and applications, 2011, Food Hydrocolloids, 25(8), 1945–1962.

[6] Adam Macierzanka, Franziska Böttger, Laura Lansonneur, Rozenn Groizard, Anne-Sophie Jean, Neil M. Rigby, Kathryn Cross, Nikolaus Wellner, Alan R. Mackie The effect of gel structure on the kinetics of simulated gastrointestinal digestion of bovine b-lactoglobulin, 2012, Food Chemistry 134 2156–2163.

[7] Qing Guo, Aiqian Ye, Mita Lad, Douglas Dalgleishb, Harjinder Singh Effect of gel structure on the gastric digestion of whey protein emulsion gels, 2014, Soft Matter, 10, 1214-1223. DOI: 10.1039/C3SM52758A

[8] Luo, Qi, Borst, Jan Willem, Westphal, Adrie H., Boom, Remko M., Janssen, Anja E.M. Pepsin diffusivity in whey protein gels and its effect on gastric digestion, 2017, Food hydrocolloids, 66, 318-325

[9] Barbé F., Ménard O., Le Gouar Y., Buffière C., Famelart M. H., Laroche B., Le Feunteun S., Dupont D., Rémond D. The heat treatment and the gelation are strong determinants of the kinetics of milk proteins digestion and of the peripheral availability of amino acids, 2013, Food Chemistry, 136, 1203-1212.

[10] *Kéra Nyemb-DiopKéra, Catherine Guérin, Stéphane Pézennec, Françoise Nau The structural properties of egg white gels impact the extent of in vitro protein digestion and the nature of peptides generated, 2016, Food Hydrocolloids Part B, 54, 315-327.*

[11] *Mauricio Opazo-Navarrete, Marte D. Altenburg, Remko M. Boom, Anja E. M. Janssen The Effect of Gel Microstructure on Simulated Gastric Digestion of Protein Gels Food Biophysics, 2018, 13:124–138 DOI:10.1007/s11483-018-9518-7*

[12] *Floury J., Bianchi T., Thévenot J., Dupont D., Jamme F., Lutton E., Panouillé M., Boué F., Le Feunteun S. Exploring the breakdown of dairy protein gels during in vitro gastric digestion using time-lapse synchrotron deep-UV fluorescence microscopy, 2017, Food Chemistry, 239, 898-910.*

[13] *Minekus M., Alminger M., Alvito P., Ballance S., Bohn T., Bourlieu C., Carrière F., Boutrou R., Corredig M., Dupont D., Dufour C., Egger L., Golding M., Karakaya S., Kirkhus B., Le Feunteun S., Lesmes U., Macierzanka A., Mackie A., Marze S., McClements D. J., Ménard O., Recio I., Santos C. N., Singh R. P., Vegarud G. E., Wickham M. S. J., Weitschies W., and Brodkorb A. A standardized static in vitro digestion method suitable for food–an international consensus, 2014, Food & Function, 5, 1113.*

[14] *Forster S., Hermsdorf N., Leube W., Schnablegger H., Regenbrecht M., Akari S., Lindner P. and Böttcher C. Fusion of Charged Block Copolymer Micelles into Toroid Networks, 1999, J. Phys. Chem., B103, 6657-6668*

[15] *Rouf C., Munch J. P., Schosseler F., Pouchelon A., Beinert G., Boué F., Bastide J., Thermal and Quenched fluctuations of polymer concentration in polydimethylsiloxane gels, 1997, Macromolecules, 30, 8344-8359..*

[16] *Sebastian Seiffert Scattering perspectives on nanostructural inhomogeneity in polymer network gels, 2017, Progress in Polymer Science 66 () 1–21*

[17] *Mitsuhiro Shibayama Spatial inhomogeneity and dynamic fluctuations of polymer gels, 1998, Macromol. Chem. Phys. 199, 1-30*

[18] *Samuel Guillot, Didier Lairez, Monique A.V. Axelos Non-self-similar aggregation of methylcellulose, 2000, J. Appl. Cryst. 33, 669-672*

[19] *Satoshi Koizumi, Michael Monkenbusch, Dieter Richter, Bela Farago Concentration Fluctuations in Polymer Gel Investigated by Neutron Scattering: Static Inhomogeneity in Swollen Gel, 2005, Journal of Chemical Physics 121(24):12721-31*

[20] *http://www.sasview.org*